\documentstyle[12pt,epsf]{article}
\catcode`\@=11 \@addtoreset{equation}{section}  %%
\renewcommand{\theequation}                     %%
         {\arabic{section}.\arabic{equation}}   %%
\title{The Quantum Relativity and Dynamical Spacetime}
\author{Peter Leifer}
\date{Or-Aqiva, Israel}
\begin{document}
\maketitle
\begin{abstract}
Quantum field theory (QFT) based on the principles of special relativity (SR) and it is in fact the \emph{kinematic theory of fields}. The root assumption is that there is ``relativistic description"  of \emph{any} isolated quantum system in the so-called class of inertial systems even if the internal interactions or self-interactions lie outside of the formal QFT itself. In such a situation we cannot be sure that the principle of relativity in the present form is universally applicable since this principle arose from the Maxwell electrodynamics. As we know Einstein was insisted to generalize this principle in the attempt to find the relativistic description of gravity. Together with this the Galileo-Newton principle of inertia was modified with essential reservations \cite{Einstein_1921,Le13,Le15,Le16,Le18}. New kind of sub-atomic interactions have definitely more complicated nature and mostly unknown laws. It is clear that the present QFT (kinematic theory of fields) may serve merely as a limit of some \emph{dynamical theory of quantum fields}.

\end{abstract}

PASC: 03.65 Pm, 03.65 Ca
\section{Introduction}
E. Schr\"odinger and A. Einstein discussed the principle difficulties concerning the notions of particles and acceleration as fundamental entities in the modern physics about 70 years ago \cite{Schr_50,Einstein_50}.
The mist of probability partly hides these fundamental difficulties. But the fundamental questions have no answers up to now \cite{Oxford_Q}.

The deep rooted assumption about existence of isolated system such as ``body" or ``particle" is very contradictable even in the framework of the elementary quantum physics since, say, a free electron has indefinitely large size due to the plane wave function and, hence, cannot be ``isolated". However, it is impossible simply discard this intuitively absurd result since the plane wave solution of the Dirac equation leads to the nicely confirmed the dispersion law on the ``mass shell". This is the simplest example of the fundamental unsolved problems related to the localization, divergences, etc.,\cite{Le13,Le16,Le18}.

The classical ``body" or ``particles" and their abstract model - ``material point" cannot be used as a primordial elements of the consistent QFT. The distance between bodies - the root mathematical discriminator of classical physics is very vague parameter in quantum theory.  The second fundamental classical primordial element is the ``force" as some external relative to the ``body" factor capable to change the body state - its coordinates and velocity.  Quantum theory hides the ``force" in a shadow since the reason of the perturbation of quantum motion is not a force but the EM potential changes momentum $\vec{p}=m\vec{v}$ in the additive manner $\vec{P}=m\vec{v}+\frac{e}{c}\vec{A}$. It turns out that
the nature of momentum and potential requires to take into account the fundamental symmetry of interaction and in fact a new geometry of spacetime as it was in the case of the Maxwell electrodynamics leading to the relativistic mechanics of Einstein.
The Scr\"odinger picture even in the case of the relativistic Dirac equation is quite acceptable on the atomic level but its continuation down to the sub-atomic (nuclear) level in very questionable.
Probably, the long living attempts to attribute the universal spacetime environment for any ``elementary" particle like electron may be a fundamentally wrong. Such embedding of particles into pseudo-Euclidian (or pseudo-Riemannian) single spacetime seems to be so obvious and absolutely natural that even some doubts look like a mockery on the common sense. Nevertheless it is worse while to take into account that our every day experience even with modern scientific background is mostly mesoscopic and just the microscopic and global picture of the world are controversial. But up to now physicist's majority assumes that the relativistic symmetry is enough restrictive (besides GR) in order to build correct QFT (in the spirit of SM and even beyond it). This is one of the most erroneous assumption in modern physics.

In order to see this more clear it is useful return to classical mechanics of Newton where physical parameters and geometrical values are perfectly separated (in QFT such separation is questionable). Say, in the second law $m\frac{d^2 \vec{x}}{dt}=\vec{F}$ acceleration is pure geometric object (vector) whereas the mass is a scalar physical parameter. All is good if one deals with pointwise particle or absolutely solid body. Then given empirically mass of the body and its shape together with applied force and moment define the acceleration and trajectory. If a body, however, is \emph{deformable} under the action of the external forces or in the case of the ``falling cat" motion, then the dynamics will be more complicated and requires essentially new gauge theory of finite deformations \cite{Littlejohn}. Applying this apparatus in QFT one shall remember that there is essential simplification in the classical case in comparison with the QFT. Namely, one may be sure that the macroscopic rods, masses, etc., move in 3D Euclidian space, whereas in the QFT case there are serious doubts in the motion of quantum states of single ``elementary" particle in the 4D pseudo-Euclidean or pseudo-Riemannian spacetime. Shortly speaking - initially one should forget about ``particle" and think about the motion of the quantum state in some functional space.

The intrinsic unification of the quantum theory and relativity is possible only on the way of the serious deviation from traditional assumptions about a priori spacetime structure and the Yang-Mills generalization of the well known $U(1)$ Abelian gauge symmetry of the classical electrodynamics. In fact, more general gauge theory should be constructed.
Formally we deal with the quantum version of the gauge theory of the deformable bodies - the gauge theory of the deformable quantum state. More physically this means that the distance between quantum states is strictly defined value whereas the distance between bodies (particle) is an approximate value, at best \cite{Le13,Le15,Le16,Le18}. Thereby, all well known solid frames and clocks even with the corrections of special relativity should be replaced by the flexible and anholonomic quantum setup. Then Yang-Mills arguments about the spacetime coordinate dependence of the gauge unitary rotations should be reversed on the  dependence of the spacetime structure on the gauge transformations of the flexible quantum setup. One needs to build ``inverse representation" of the unitary transformations by the intrinsic dynamical spacetime transformations.

In order to build such DST one needs the general footing for gauge fields and for ``matter fields". Only fundamental pure quantum degrees of freedom like spin, charge, hyper-charges, etc., obey this requirement. One may assume that they correspond some fundamental quantum motions in the manifold of the UQS's. Then ``elementary particles" will be represented as a dynamical process keeping non-linear coherent superposition of these fundamental quantum motions.
One should, however, distinct the ``total quantum state" (cum location) as an analog of the spatial coordinates of the system of material points with their ``orientation coordinates", and the ``unlocated quantum state" of the quantum degrees of freedom (QDF's) as an analog of the ``unlocated shape coordinates".

The spacetime and its transformations will be built ``from inside"  of the ``elementary" particle due to the separation of the group $L_+^{\uparrow}$ from the $G=SU(N)$ acting on the quantum state space of rays $CP(N-1)$ by the diffeomorphic coset transformations $G/H =SU(N)/S[U(1) \times U(N-1)]= CP(N-1) $.

\section{Manifold of the Quantum Degrees of Freedom}
The motions of the quantum degrees of freedom (QDF's) under the unitary transformations comprises the manifold of the unlocated quantum states (UQS's). These ``elementary" motions replace ``elementary particles" of the Standard Model. Its soliton-like excitations then realized as known ``elementary particles". The intrinsic ``unitary field" acting without super-selection rule continuously splits the multiplete of the spin, charge, hypercharge, etc., into zones.  QDF's acts as unified ``chiral" field whose dynamics will be discussed properly.

The fundamental quantum degrees of freedom like spin, charge, hyper-charges, etc., are common for gauge and matter fields. These fundamental quantum motions take the place in the manifold of the UQS's which described by the rays of states $|\psi> \in C^N$ of the ``unitary spin" $S: 2S+1=N$ . Physics requires to use in this background the local coordinates of UQS's and the state-dependent generators of the unitary group $G=SU(N)$ \cite{Le97,Le98,Le04}. This nonlinear representation of the $SU(N)$ group on the coset manifold $G/H=SU(N)/S[U(1) \times U(N-1)]=CP(N-1)$ is primary and this is independent on the spacetime manifold. The last one should be introduced in a special section of the fiber bundle over $CP(N-1)$ \cite{Le97,Le98,Le99,Le11,Le13,Le15,Le16,Le18}. The breakdown of the global $SU(N)$ symmetry down to the isotropy subgroup $H_{|\psi>}=U(1) \times U(N-1)$ of the some quantum state $|\psi>$ has natural geometric counterpart in $CP(N-1)$.

The coset manifold $G/H_{|\psi>}=SU(N)/S[U(1) \times U(N-1)]=CP(N-1)$ contains locally unitary transformations \emph{deforming} ``initial" quantum state $|\psi>$. This means that $CP(N-1)$ contains physically distinguishable, ``deformed" quantum states. Thereby the unitary transformations from $G=SU(N)$ of the basis in the Hilbert space may be identified with the unitary state-dependent gauge field $U(|\psi>)$ that may be represented by the $N^2-1$ unitary generators as functions of the local projective coordinates $(\pi^1,...,\pi^{N-1})$ \cite{Le13}. This manifold resembles the ``shape space" of the deformable body \cite{Littlejohn,Le13,Le15,Le16,Le18}. But now it is the manifold of the deformed physically distinguishable UQS's, i.e. the geometric, invariant counterpart of the quantum interaction or self-interaction. Then the classical acceleration is merely an ``external" consequence of this complicated quantum dynamics in the some section of the frame fiber bundle over $CP(N-1)$. The local dynamical variables (LDV's) are new essential elements of the new quantum dynamics \cite{Le04}. They should be expressed in terms of the local coordinates $\pi^k$ of UQS's. Thereby they
will live in the geometry of $CP(N-1)$ with the Fubini-Study metric tensor
\begin{equation}
G_{ik^*} = (1/{\kappa})[(1+ \sum |\pi^s|^2) \delta_{ik}- \pi^{i^*} \pi^k](1+
\sum |\pi^s|^2)^{-2},
\end{equation}\label{2}
where $\kappa$ is holomorphic sectional curvature of the $CP(N-1)$ \cite{KN}.
The contra-variant metric tensor field
\begin{equation}
G^{ik^*} =\kappa (\delta^{ik} + \pi^{i} \pi^{k*})(1+
\sum |\pi^s|^2),
\end{equation}\label{3}
is inverse to the $G_{ik^*}$ thereby
\begin{equation}
G_{ik^*}G^{i^*q} = \delta_{k}^q.
\end{equation}\label{3}
The affine connection agrees with the Fubini-Study metric is as follows
\begin{eqnarray}
\Gamma^i_{mn} = \frac{1}{2}G^{ip^*} (\frac{\partial
G_{mp^*}}{\partial \pi^n} + \frac{\partial G_{p^*n}}{\partial
\pi^m}) = -  \frac{\delta^i_m \pi^{n^*} + \delta^i_n \pi^{m^*}}{1+
\sum |\pi^s|^2}.
\end{eqnarray}\label{3}
The curvature tensor of Riemann in holonomic basis is proportional to the constant section curvature since
\begin{eqnarray}
R^i_{klm^*} = \kappa^2 (\delta^i_{l}G_{km^*}+\delta^i_{k}G_{lm^*})
\end{eqnarray}
\cite{KN}.
The flexible quantum setup inherently connected with local projective coordinates will be built from so-called LDV's \cite{Le04}. These LDV's realize
a non-linear representation of the unitary global $SU(N)$ group in
the Hilbert state space $C^N$. Namely, $N^2-1$ generators of $G =
SU(N)$ may be divided in accordance with the Cartan decomposition:
$[B,B] \in H, [B,H] \in B, [B,B] \in H$. The $(N-1)^2$ generators
\begin{eqnarray}
\Phi_h^i \frac{\partial}{\partial \pi^i}+c.c. \in H,\quad 1 \le h
\le (N-1)^2
\end{eqnarray}\label{4}
of the isotropy group $H = U(1)\times U(N-1)$ of the ray (Cartan
sub-algebra) and $2(N-1)$ generators
\begin{eqnarray}
\Phi_b^i \frac{\partial}{\partial \pi^i} + c.c. \in B, \quad 1 \le b
\le 2(N-1)
\end{eqnarray}\label{5}
are the coset $G/H = SU(N)/S[U(1) \times U(N-1)]$ generators
realizing the breakdown of the $G = SU(N)$ symmetry.
Notice, the partial derivatives are defined here as usual: $\frac{\partial }{\partial \pi^i} = \frac{1}{2}
(\frac{\partial }{\partial \Re{\pi^i}} - i \frac{\partial }{\partial
\Im{\pi^i}})$ and $\frac{\partial }{\partial \pi^{*i}} = \frac{1}{2}
(\frac{\partial }{\partial \Re{\pi^i}} + i \frac{\partial }{\partial
\Im{\pi^i}})$.

Here $\Phi^i_{\sigma}, \quad 1 \le \sigma \le N^2-1 $
are the coefficient functions of the generators of the non-linear
$SU(N)$ realization. They give the infinitesimal shift of the
$i$-component of the generalized coherent state driven by the $\sigma$-component
of the unitary  field $\exp(i\epsilon \lambda_{\sigma})$ rotating by the
generators of $Alg SU(N)$ and they are defined as follows:
\begin{equation}
\Phi_{\sigma}^i = \lim_{\epsilon \to 0} \epsilon^{-1}
\biggl\{\frac{[\exp(i\epsilon \lambda_{\sigma})]_m^i \psi^m}{[\exp(i
\epsilon \lambda_{\sigma})]_m^j \psi^m }-\frac{\psi^i}{\psi^j} \biggr\}=
\lim_{\epsilon \to 0} \epsilon^{-1} \{ \pi^i(\epsilon
\lambda_{\sigma}) -\pi^i \},
\end{equation}\label{6}
\cite{Le13}.

\section{Quantum relativity}
The principle of Quantum Relativity (QR) was initially called "super-relativity" \cite{Le97,Le98}) assumes the invariance of physical properties of ``quantum particles" i.e. their quantum numbers like mass, spin, charge, etc.  Such invariance may be lurked, say, behind two amplitudes $|\Psi_1>, |\Psi_2>$ in two different quantum setups $S_1$ and $S_2$.  The invariant content of these properties will be discussed here under the infinitesimal variation of the ``flexible quantum setup" described by the amplitudes $|\Psi(\pi,P)>$ due to a small variation of the boson electromagnetic-like field $P^{\alpha}(\pi)$ treated as \emph{the set of the scalar functions} relative ${\pi^i}$ coordinates in $CP(N-1)$. The DST dependence of $P^{\alpha}(\pi)$ will be established after the separation of the shifts, boosts and rotations in the manifold of the $SU(N)$ generators.

The mathematical formulation of the QR principle is based on the \emph{similarity} of any physical systems (``setup", if somebody wants) which are built on the ``elementary" particles. This similarity is obvious only on the level of the pure quantum degrees of freedom of quantum particles. Therefore, all ``external" details of the ``setup" should be discarded as non-essential and only the relations of components of the ``unitary spin" like $(\pi^1=\frac{\psi^2}{\psi^1},...,\pi^{N-1}=\frac{\psi^{N}}{\psi^1})$  should be taken into account. These relations will be taken as the local projective coordinates in the complex projective Hilbert space $CP(N-1)$. One may think about these coordinates as parameters of the ``shape of quantum particle" in the spirit of the \cite{Littlejohn}. This ``shape" is the unlocated quantum state (UQS) of the ``unitary spin" $S=\frac{N-1}{2}$. These coordinates are analog of an \emph{angle} in the trigonometry that is the invariant characteristic of all similar triangles. Thereby, the coefficients functions $\Phi^i_{\alpha}$ of the generators of $SU(N)$ defined as the Lie derivative of the local projective coordinates $\pi^i$ under the infinitesimal unitary variation of the appropriate parameter $\epsilon$ in (2.8).

\section{Separation of the Poincar\'e generators from the $Alg SU(4)$ and the dynamical spacetime }
The old problem of the accelerated charged particle is an acute challenge for QFT, high energy physics, and for the theory of elementary particles. There is an interesting attempt to solve this problem in the spirit of my concept of the deformation of UQS \cite{Penheiro}. Namely, the ``backreaction of space" clearly close to the DST concept \cite{Le11,Le13,Le15,Le16,FQXi_2016}.

It worse while to say two words about so-called Quantum Potential \cite{Bohm}.
Since in the depth of the ``elementary" particle there is no the ordinary ``physical" spacetime and DST has state-dependent nature, the ordinary gradient of the momentum in respect of pseudo-Euclidian coordinates has no sense. But this concept is very prolific for the intermolecular interactions that nicely demonstrated in the Complex Mechanics of Yang and its applications \cite{Yang1}. Furthermore, the complexification of the spacetime coordinates and momenta is the step in the correct direction with the reservation about the functional nature of these coordinates: they are functions in some gauge ``sheet" in the Hilbert state space.

The metric of the DST is state-dependent that may be demonstrated directly by the calculations of the square of the speed velocity $\frac{dS^2}{d \tau^2}$ of the geodesic distance in $CP(3)$ \cite{Le18}. ``Diffusion" of the mass-shell is evident here but the scale of such diffusion is unknown since the value of the sectional curvature $\kappa$ included in $G_{ik^*}$ is a free parameter up to now. It is closely connected with the non-separability
of the inertial mass $m$ from the acceleration $\frac{d^2 x}{dt^2}$ in the Newton's expression for the force $\frac{dp}{dt} = m \frac{d^2 x}{dt^2} = -\frac{\partial V}{\partial x}$. \emph{The speed of the momentum variation will be treat now as field equation with TWS solution instead of the equation for trajectory of the point-wise particle.}

In the case of the extended quantum electron \cite{Le11} one deals with the dynamical reconstruction of the $G=SU(4)$ symmetry of the UQS down to the $H=S[U(1) \times U(3)]$ of the unitary rotation of the whole $CP(3)$ geodesics.
This geodesic carriers unperturbed dynamics of the self-interacting ``free" electron. This internal dynamics of the spin/charge degrees of freedom in $CP(3)$ cannot be directly connected with spacetime since this is ``unlocated". Its spacetime coordinates are absent. Furthermore, we deal not with the ``events" as points of the Minkowski spacetime but with a dynamical process. There is an analogy with the MRI method  \cite{Hornak} of the coordinate prescription for voxeles in the (field-of-view) FOV and the ``inverse representation" of the UQS's motion in $CP(N-1)$ by the field dynamics in the DST \cite{Le13,Le15,Le16}. The MRI coordinate introduction is based on the known dependence of the frequency shift on the distance in the gradient magnetic field. Then known field amplitude and frequency gives the distance according very simple formula $\delta \overrightarrow{x} = k \nabla B$. In our case the spacetime distribution of the gauge field and total wave function are unknown and they should be found.

I will introduce the definition of the Poincar\'e generators in the local DST as the special linear combinations of the Lie derivatives of the local projective coordinates $(\pi^1,\pi^2,\pi^3)$ in directions given by the Dirac matrices in the Weyl representation. This construction is most transparent for the fundamental fermion like the electron. More general case of higher dimension should be discussed elsewhere. For this aim I will use the following set of the Dirac matrices
\begin{eqnarray}\label{43}
\gamma_t=\left( \begin {array}{cccc}
0&0&1&0 \cr
0&0&0&1 \cr
1&0&0&0 \cr
0&1&0&0
\end {array} \right),
\gamma_1=-i \sigma_1=\left( \begin {array}{cccc}
0&0&0&-1 \cr
0&0&-1&0 \cr
0&1&0&0 \cr
1&0&0&0
\end {array} \right), \cr
\gamma_2=-i \sigma_2=\left( \begin {array}{cccc}
0&0&0&i \cr
0&0&-i&0 \cr
0&-i&0&0 \cr
i&0&0&0
\end {array} \right),
\gamma_3=-i \sigma_3=\left( \begin {array}{cccc}
0&0&-1&0 \cr
0&0&0&1 \cr
1&0&0&0 \cr
0&-1&0&0
\end {array} \right).
\end{eqnarray}
Then the corresponding coefficients of the $SU(4)$ generators will be calculated according to the equation
\begin{equation}
\Phi_{\mu}^i = \lim_{\epsilon \to 0} \epsilon^{-1}
\biggl\{\frac{[\exp(i\epsilon \gamma_{\mu})]_m^i \psi^m}{[\exp(i
\epsilon \gamma_{\mu})]_m^j \psi^m }-\frac{\psi^i}{\psi^j} \biggr\}=
\lim_{\epsilon \to 0} \epsilon^{-1} \{ \pi^i(\epsilon
\gamma_{\mu}) -\pi^i \},
\end{equation}\label{6}
\cite{Le13}
that gives
\begin{eqnarray}
\Phi_{0}^1(\gamma_{t})&=&i(\pi^3-\pi^1 \pi^2), \quad \Phi_{0}^2(\gamma_{t})=i(1-(\pi^2)^2),
\quad \Phi_{0}^3(\gamma_{t})=i(\pi^1-\pi^2 \pi^3); \cr
\Phi_{1}^1(\gamma_{1})&=&-i(\pi^2 -\pi^1 \pi^3),
\quad \Phi_{1}^2(\gamma_{1})=-i(-\pi^1 -\pi^2 \pi^3),
\quad \Phi_{1}^3(\gamma_{1})=-i(-1 -(\pi^3)^2); \cr
\Phi_{2}^1(\gamma_{2})&=&-i(i(\pi^2 +\pi^1 \pi^3)),
\quad \Phi_{2}^2(\gamma_{2})=-i(i(\pi^1 +\pi^2 \pi^3)),
\quad \Phi_{2}^3(\gamma_{2})=-i(i(-1 +(\pi^3)^2)); \cr
\Phi_{3}^1(\gamma_{3})&=&-i(-\pi^3 -\pi^1 \pi^2),
\quad \Phi_{3}^2(\gamma_{3})=-i(-1 -(\pi^2)^2),
\Phi_{3}^3(\gamma_{3})=-i(\pi^1 -\pi^2 \pi^3).
\end{eqnarray}\label{15}
Such choice of the vector fields lead to the ``imaginary" basic in local DST which conserves $4D$ Eucledian geometry along geodesic in $CP(3)$ for real four vectors $(p^0,p^1,p^2,p^3)$ and correspondingly $4D$ pseudo-Eucledian geometry for four vectors $(ip^0,p^1,p^2,p^3)$.

The complex DST ``tangent vector" in $\mu$ direction defines the four complex shifts in DST that will be introduced as follows:
\begin{eqnarray}
\frac{\partial }{\partial x^{\mu}} = \Phi_{\mu}^i \frac{\partial }{\partial \pi^i}
\end{eqnarray}\label{}
for $0\leq \mu \leq 3$. In fact one may define the similar ``tangent vector" in $\alpha$ direction
\begin{eqnarray}
\frac{\partial }{\partial x^{\alpha}} = \Phi_{\alpha}^i \frac{\partial }{\partial \pi^i}
\end{eqnarray}\label{}
for $1 \leq \alpha \leq 15$ in the space $R^{15}$ of the adjoint representation of the $SU(4)$. Thereby, the DST cannot be treated as the ``space of events". It is rather 10-dimension subspace of the adjoint representation of the $SU(4)$. The quantum operator of the energy-momentum will be expressed as the shift operator
\begin{eqnarray}
\vec{P}_{\mu}=i\hbar \frac{\partial }{\partial x^{\mu}} = i\frac{\hbar}{L}  \Phi_{\mu}^i \frac{\partial }{\partial \pi^i}.
\end{eqnarray}\label{}
Now one may introduce six generators of the boosts and rotations started from the well known definitions in terms of Dirac matrices \cite{Feynman}.

\begin{eqnarray}\label{43}
B_x=(i/2)\gamma_t \gamma_x=(i/2) \left( \begin {array}{cccc}
0&1&0&0 \cr
1&0&0&0 \cr
0&0&0&-1 \cr
0&0&-1&0
\end {array} \right),\cr
B_y=(i/2) \gamma_t \gamma_y=(i/2)\left( \begin {array}{cccc}
0&i&0&0 \cr
-i&0&0&0 \cr
0&0&0&-i \cr
0&0&i&0
\end {array} \right), \cr
B_z=(i/2) \gamma_t \gamma_z=(i/2)\left( \begin {array}{cccc}
1&0&0&0 \cr
0&-1&0&0 \cr
0&0&-1&0 \cr
0&0&0&1
\end {array} \right),\cr
R_x=(i/2) \gamma_y \gamma_z= (i/2) \left( \begin {array}{cccc}
0&i&0&0 \cr
i&0&0&0 \cr
0&0&0&i \cr
0&0&i&0
\end {array} \right),\cr
R_y=(i/2) \gamma_z \gamma_x= (i/2) \left( \begin {array}{cccc}
0&-1&0&0 \cr
1&0&0&0 \cr
0&0&0&-1 \cr
0&0&1&0
\end {array} \right),\cr
R_z=(i/2) \gamma_x \gamma_y= (i/2) \left( \begin {array}{cccc}
i&0&0&0 \cr
0&-i&0&0 \cr
0&0&i&0 \cr
0&0&0&-i
\end {array} \right).
\end{eqnarray}
Using the modified definition (2.8) one may find the corresponding coefficient functions of the vector fields of the Lorentz generators for boosts
\begin{equation}
\Phi^i(B_\alpha) = \lim_{\epsilon \to 0} \epsilon^{-1}
\biggl\{\frac{[\exp(\epsilon B_{\alpha})]_m^i \psi^m}{[\exp(
\epsilon B_{\alpha})]_m^j \psi^m }-\frac{\psi^i}{\psi^j} \biggr\}=
\lim_{\epsilon \to 0} \epsilon^{-1} \{ \pi^i(\epsilon
B_{\alpha}) -\pi^i \},
\end{equation}\label{6}
\begin{eqnarray}
\Phi^1(B_x)=\frac{1}{2}(1-(\pi^1)^2),
\Phi^2(B_x)=\frac{-1}{2}(\pi^3+\pi^1\pi^2),
\Phi^3(B_x)=\frac{-1}{2}(\pi^2+\pi^1\pi^3),\cr
\Phi^1(B_y)=-\frac{i}{2}(1+(\pi^1)^2),
\Phi^2(B_y)=-\frac{i}{2}(\pi^3+\pi^1\pi^2),
\Phi^3(B_y)=\frac{i}{2}(\pi^2-\pi^1\pi^3),\cr
\Phi^1(B_z)=-\pi^1,
\Phi^2(B_z)=-\pi^2,
\Phi^3(B_z)=0,
\end{eqnarray}
and rotations
\begin{equation}
\Phi^i(R_\alpha) = \lim_{\epsilon \to 0} \epsilon^{-1}
\biggl\{\frac{[\exp(\epsilon R_{\alpha})]_m^i \psi^m}{[\exp(
\epsilon R_{\alpha})]_m^j \psi^m }-\frac{\psi^i}{\psi^j} \biggr\}=
\lim_{\epsilon \to 0} \epsilon^{-1} \{ \pi^i(\epsilon
R_{\alpha}) -\pi^i \},
\end{equation}\label{6}

\begin{eqnarray}
\Phi^1(R_x)=\frac{i}{2}(1-(\pi^1)^2),
\Phi^2(R_x)=\frac{i}{2}(\pi^3-\pi^1\pi^2),
\Phi^3(R_x)=\frac{i}{2}(\pi^2-\pi^1\pi^3),\cr
\Phi^1(R_y)=\frac{1}{2}(1+(\pi^1)^2),
\Phi^2(R_y)=-\frac{1}{2}(\pi^3-\pi^1\pi^2),
\Phi^3(R_y)=\frac{1}{2}(\pi^2+\pi^1\pi^3),\cr
\Phi^1(R_z)=-i\pi^1,
\Phi^2(R_z)=0,
\Phi^3(R_z)=-i\pi^3,
\end{eqnarray}
Thereby, the eight $\lambda$-matrices  $(\lambda_4,\lambda_{11}),(\lambda_2,\lambda_{14}),(\lambda_1,\lambda_{13}),(\lambda_5,\lambda_{12})$ of the $AlgSU(4)$ were involved in the definition of the shift vector fields. It is easy to see that additional diagonal matrices $,(\lambda_{3}),(\lambda_{8}),(\lambda_{15})$ must be involved into the boosts and rotations definitions whereas four $\lambda$-matrices $(\lambda_6,\lambda_{7}),(\lambda_9,\lambda_{10})$ mixing electron-positron states with opposite spins all together with the eight $\lambda$-matrices comprise of the full set of the fifteenth matrices of the $Alg SU(4)$.

Then the three generators
\begin{eqnarray}
\vec{B}_{\alpha}=  \Phi^i(B_{\alpha}) \frac{\partial }{\partial \pi^i}
\end{eqnarray}\label{}
define the boosts and three generators
\begin{eqnarray}
\vec{R}_{\alpha}=  \Phi^i(R_{\alpha}) \frac{\partial }{\partial \pi^i}.
\end{eqnarray}\label{}
define the rotations.
The commutators of these vector fields is as follows.
\begin{eqnarray}
[P_0,P_1]=(\xi^1 = 2(1+ (\pi^1)^2 , \xi^2= -2(\pi^1 \pi^2 + \pi^3), \xi^3 = -2(\pi^1 \pi^3 + \pi^2),\cr
[P_0,P_2]=(\xi^1=2i(1+(\pi^1)^2), \xi^2=2i(\pi^1+\pi^2\pi^3),\xi^3=2i(\pi^1\pi^3-\pi^2)),\cr
[P_0,P_3]=(\xi^1=-4\pi^1,\xi^2=-4\pi^2,\xi^3=0),\cr
[P_3,P_1]=(\xi^1=2(1+(\pi^1)^2),\xi^2=2( \pi^1 \pi^2 - \pi^3),\xi^3=2(\pi^1\pi^3+\pi^2)),\cr
[P_2,P_1]=(\xi^1=-4i\pi^1,\xi^2=0,\xi^3=-4i\pi^3),\cr
[P_3,P_2]=(\xi^1=2i(1-(\pi^1)^2),\xi^2=2i(-\pi^1 \pi^2 + \pi^3),\xi^3=2i(-\pi^1\pi^3+\pi^2)),\cr
[B_1,B_2]=(\xi^1=-i\pi^1,\xi^2=0,\xi^3=-i\pi^3),\cr
[B_1,B_3]=(\xi^1=-(1+(\pi^1)^2)/2, \xi^2=(\pi^3-\pi^1\pi^2)/2,\xi^3=-(\pi^1\pi^3+\pi^2)/2),\cr
[B_3,B_2]=(\xi^1=-i(1+(\pi^1)^2)/2, \xi^2=-i(\pi^3-\pi^1\pi^2)/2,\xi^3=-i(\pi^1\pi^3+\pi^2)/2),\cr
[R_1,R_2]=(\xi^1=i\pi^1,\xi^2=0,\xi^3=i\pi^3),\cr
[R_3,R_2]=(\xi^1=i(1-(\pi^1)^2)/2, \xi^2=i(\pi^3-\pi^1\pi^2)/2,\xi^3=i(-\pi^1\pi^3+\pi^2)/2),\cr
[R_1,R_3]=(\xi^1=(1+(\pi^1)^2)/2, \xi^2=(-\pi^3+\pi^1\pi^2)/2,\xi^3=(\pi^1\pi^3+\pi^2)/2),\cr
[B_1,R_1]=(\xi^1=0,\xi^2=0,\xi^3=0),\cr
[B_1,R_2]=(\xi^1=\pi^1,\xi^2=\pi^2),\xi^3=0,\cr
[B_1,B_3]=(\xi^1=-i(1+(\pi^1)^2)/2, \xi^2=-i(\pi^3+\pi^1\pi^2)/2,\xi^3=i(-\pi^1\pi^3+\pi^2)/2),\cr
[B_2,R_1]=(\xi^1=-\pi^1,\xi^2=-\pi^2,\xi^3=0),\cr
[B_2,R_2]=(\xi^1=0,\xi^2=0,\xi^3=0),\cr
[B_2,R_3]=(\xi^1=-(1-(\pi^1)^2)/2, \xi^2=(\pi^3+\pi^1\pi^2)/2,\xi^3=(\pi^1\pi^3+\pi^2)/2),\cr
[B_3,R_1]=(\xi^1=i(1+(\pi^1)^2)/2, \xi^2=i(\pi^3+\pi^1\pi^2)/2,\xi^3=i(\pi^1\pi^3-\pi^2)/2),\cr
[B_3,R_2]=(\xi^1=(1-(\pi^1)^2)/2, \xi^2=-(\pi^3+\pi^1\pi^2)/2,\xi^3=-(\pi^1\pi^3+\pi^2)/2),\cr
[B_3,R_3]=(\xi^1=0,\xi^2=0,\xi^3=0),\cr
[P_0,R_1]=(\xi^1=0,\xi^2=0,\xi^3=0),\cr
[P_0,R_2]=(\xi^1=0,\xi^2=0,\xi^3=0),\cr
[P_0,R_3]=(\xi^1=0,\xi^2=0,\xi^3=0),\cr
[P_0,B_1]=(\xi^1 = i(-\pi^1 \pi^3 + \pi^2) , \xi^2= -i(\pi^3 \pi^2 + \pi^1), \xi^3 =-i(1+ (\pi^3)^2),\cr
[P_0,B_2]=(\xi^1 = (\pi^1 \pi^3 + \pi^2) , \xi^2= \pi^3 \pi^2 + \pi^1, \xi^3 =(-1+ (\pi^3)^2),\cr
[P_0,B_3]=(\xi^1 = -i(\pi^1 \pi^2 + \pi^3) , \xi^2= -i(1 + (\pi^2)^2), \xi^3 =i(\pi^1 -\pi^2 \pi^3),\cr
[P_1,R_1]=(\xi^1=0,\xi^2=0,\xi^3=0),\cr
[P_1,R_2]=(\xi^1 = -i(\pi^1 \pi^2 + \pi^3) , \xi^2= -i(1 + (\pi^2)^2), \xi^3 =i(\pi^1 -\pi^2 \pi^3),\cr
[P_1,R_3]=(\xi^1 = -(\pi^1 \pi^3 + \pi^2) ,\xi^2 =-(\pi^1 +\pi^2 \pi^3), \xi^3= (1 - (\pi^3)^2),\cr
[P_1,B_1]=(\xi^1 = -i(\pi^1 \pi^2 - \pi^3) ,\xi^2 =-i(1 - (\pi^2)^2),\xi^3=-i(\pi^1 -\pi^2 \pi^3),\cr
[P_1,B_2]=(\xi^1=0,\xi^2=0,\xi^3=0),\cr
[P_1,B_3]=(\xi^1=0,\xi^2=0,\xi^3=0),\cr
[P_2,R_1]=(\xi^1 = -i(\pi^1 \pi^2 + \pi^3) ,\xi^2 =-i(1 + (\pi^2)^2) ,\xi^3=i(\pi^1-\pi^2 \pi^3),\cr
[P_2,R_2]=(\xi^1=0,\xi^2=0,\xi^3=0),\cr
[P_2,R_3]=(\xi^1 = -i(-\pi^1 \pi^2 + \pi^2) ,\xi^2 =i(\pi^1+\pi^2 \pi^3),\xi^3=i(1 + (\pi^3)^2),\cr
[P_2,B_1]=(\xi^1=0,\xi^2=0,\xi^3=0),\cr
[P_2,B_2]=(\xi^1 = i(-\pi^1 \pi^2 + \pi^3) ,\xi^2 =i(1 - (\pi^2)^2) ,\xi^3=i(\pi^1-\pi^2 \pi^3)),\cr
[P_2,B_3]=(\xi^1=0,\xi^2=0,\xi^3=0),\cr
[P_3,R_1]=(\xi^1 = \pi^1 \pi^3 + \pi^2 ,\xi^2 =\pi^1+\pi^2 \pi^3 ,\xi^3=-1 + (\pi^3)^2),\cr
[P_3,R_2]=(\xi^1 = -i(-\pi^1 \pi^3 + \pi^2) ,\xi^2 =i(\pi^1+\pi^2 \pi^3) ,\xi^3=i(1 + (\pi^3)^2),\cr
[P_3,R_3]=(\xi^1=0,\xi^2=0,\xi^3=0),\cr
[P_3,B_1]=(\xi^1=0,\xi^2=0,\xi^3=0),\cr
[P_3,B_2]=(\xi^1=0,\xi^2=0,\xi^3=0),\cr
[P_3,R_3]=(\xi^1 = i(\pi^1 \pi^2 - \pi^3),\xi^2 =i((\pi^2)^2-1),\xi^3=i(\pi^2 \pi^3-\pi^1).
\end{eqnarray}

\section{The Hilbert space of the total quantum states and its gauge ``sheets" fibration} De Broglie-Schr\"odinger corpuscle-wave duality establishes the relation between the Newton-Euler-Hamilton ODE's and the Schr\"odinger-Dirac PDE wave equations. This relation  corresponds to the statistic ensemble of quasi-classical particles. There is, however, more general type of the ``corpuscle-wave duality" in the class of the quasi-linear PDE's of the first order where the field of the velocities naturally connected with characteristic curves serving trajectories for ``particles" of some medium. I propose the quasi-linear PDE whose characteristic correspond to the trajectory of a single quantum electron.

Dynamics of UQS's in the base manifold $CP(N-1)$ serves as the ``master" rules for quantum motions.  This stems from de Broglie idea of the periodicity of some process with the frequency $\omega =\frac{mc^2}{\hbar}$ in the energy parcel \cite{DB}. I formulate the following requirement: the projection of the trajectory of a single quantum particle onto $CP(N-1)$ should be a geodesic since all geodesics in $CP(N-1)$ are closed that provide the periodicity by the natural manner. This condition formulates the gauge restriction on the divergency of energy-momentum field functions in DST.

The relativistic Hamiltonian vector field
\begin{eqnarray}
\vec{H}=  c[P^{\mu} \Phi_{\mu}^i +K^{\alpha}\Phi^i(B_{\alpha})
 +M^{\alpha}\Phi^i(R_{\alpha})] \frac{\partial }{\partial \pi^i} + c.c.
\end{eqnarray}\label{}
will be used for the eigen-value problem in terms of the PDE for the total wave function.
Then the speed of the UQS components should be satisfied the following
equation of characteristics
\begin{eqnarray}
 \frac{d \pi^i}{d \tau} =  \frac{c}{\hbar}[P^{\mu} \Phi_{\mu}^i+K^{\alpha}\Phi^i(B_{\alpha})
 +M^{\alpha}\Phi^i(R_{\alpha})]
\end{eqnarray}\label{}
where $\tau$ is the \emph{quantum elapsed time counted from the start of the entanglement process}.
The Hamiltonian vector field leads to the quasi-linear PDE ``Schr\"odinger equation"
\begin{eqnarray}\label{43}
i\hbar \frac{d \Psi(\pi,q,p)}{d\tau} = cP^{\alpha}\Phi_{\alpha}^i\frac{\partial \Psi(\pi,q,p)}{\partial \pi^i} + c.c.= E[\Psi(\pi,q,p)] \Psi(\pi,q,p),
\end{eqnarray}
where the coordinates $(p,q)$ correspond to the shifts, rotations, boosts and gauge parameters of the local DST, and $E[\Psi(\pi,q,p)]$ is a functional of the total quantum state.
One should remember that the normal vector to the solution $ \Psi(\pi,q,p)$
\begin{eqnarray}
\vec{N}= (\frac{\partial \Psi(\pi,q,p)}{\partial \pi^1}, \frac{\partial \Psi(\pi,q,p)}{\partial \pi^2},\frac{\partial \Psi(\pi,q,p)}{\partial \pi^3},-1)
\end{eqnarray}
and the tangent vector
\begin{eqnarray}
\vec{T}= (cP^{\alpha}\Phi_{\alpha}^1,   cP^{\alpha}\Phi_{\alpha}^2,cP^{\alpha}\Phi_{\alpha}^3,i\hbar \frac{d \Psi(\pi,q,p)}{d\tau} ) \cr
=(cP^{\alpha}\Phi_{\alpha}^1,   cP^{\alpha}\Phi_{\alpha}^2,cP^{\alpha}\Phi_{\alpha}^3,E \Psi(\pi,q,p)),
\end{eqnarray}
so that the ``Schr\"odinger equation" may be written as the scalar product $(\vec{N}\vec{T})=0$, belong to some infinite dimensional functional space like the space of the complex value analytical functions.

In order to find physically acceptable solutions of this equation one needs to put the gauge and the ``border" restrictions on meanwhile undefined functions $P^{\alpha}$.
Our requirement tells that the projection of the trajectory of a single quantum particle onto $CP(N-1)$ should be a geodesic. Hence, the covariant derivative in the sense of the Fubini-Study metric of the velocity of UQS $\frac{d \pi^i}{d \tau}$ should be zero
\begin{eqnarray}\label{43}
( P^{\alpha}\Phi_{\alpha}^i)_{;k} = \frac{\partial P^{\alpha}}{\partial \pi^k}\Phi_{\alpha}^i + P^{\alpha} (\frac{\partial \Phi_{\alpha}^i}{\partial \pi^k}+
\Gamma^i_{kl} \Phi_{\alpha}^l) = 0.
\end{eqnarray}
One sees that the dynamical system for non-linear field momentum is self-consistent since the speed of the traversing the geodesic in $CP(N-1)$ is not a constant but a variable value ``modulated" by the field coefficients $P^{\alpha}$.

Let me take initially only the shifts in DST without rotations and boosts. Then in the equation (5.6) one will have the summation only of four terms
\begin{eqnarray}\label{43}
( P^{\mu}\Phi_{\mu}^i)_{;k} = \frac{\partial P^{\mu}}{\partial \pi^k}\Phi_{\mu}^i + P^{\mu} (\frac{\partial \Phi_{\mu}^i}{\partial \pi^k}+
\Gamma^i_{kl} \Phi_{\mu}^l) = 0.
\end{eqnarray}
In order to get the field equations in DST I use the definition of the DST derivative. Thus one may rewrite this equations for $k=i$ as follows
\begin{eqnarray}\label{43}
 \frac{\partial P^{\mu}}{\partial x^{\mu}} + P^{\mu} (\frac{\partial \Phi_{\mu}^i}{\partial \pi^i}+
\Gamma^i_{il} \Phi_{\mu}^l) = 0.
\end{eqnarray}
Thus one has the gauge restriction in the form of the field equation. For the parallel transported $\Phi_{\mu}^i$ this gauge restriction coincides with the ordinary Lorentz gauge. This linear PDE has the traveling wave solutions (TWS), say, in the form
\begin{eqnarray}
 P^{\mu} = K^{\mu}+ A^{\mu}F(\Phi_{\mu}^i) \tanh(C_0+C_1x+C_2y+C_3z+C_4t) \cr + B^{\mu}G(\Phi_{\mu}^i)\tanh(C_0+C_1x+C_2y+C_3z+C_4t)^2 + H^{\mu}(\Phi_{\mu}^i).
\end{eqnarray}
Such solutions realize the state-dependent gauge conditions on the energy-momentum (potentials) and show that in the given definition of the DST coordinates $x^{\mu}$ the complicated highly nonlinear field equations (5.7) transform into the linear PDE's (5.8) with soliton-like solution (5.9) or within more wide class of TWS's.

In general case of the full Poincar\'e motions in $10D$ DST one has correspondingly
\begin{eqnarray}\label{43}
 \frac{\partial P^{\mu}}{\partial x^{\mu}} + P^{\mu} (\frac{\partial \Phi_{\mu}^i}{\partial \pi^i}+
\Gamma^i_{il} \Phi_{\mu}^l)\cr +
 \frac{\partial K^{\alpha}}{\partial u^{\alpha}} + K^{\alpha} (\frac{\partial \Phi(B_{\alpha}^i)}{\partial \pi^i}+
\Gamma^i_{il} \Phi(B_{\alpha}^l) \cr+
\frac{\partial M^{\alpha}}{\partial {\omega}^{\alpha}} + M^{\alpha} (\frac{\partial \Phi(R_{\alpha}^i)}{\partial \pi^i}+
\Gamma^i_{il} \Phi(R_{\alpha}^l)) = 0.
\end{eqnarray}
with more complicated but similar TWS solutions. Nevertheless since each such solution contains the $\pi^i$ coordinates only in the rational manner the PDE's for the parallel transport condition (5.6) will be pure algebraic. Therefore, one has the field of the energy-momentum in the local $10D$ DST as the functions of $\pi^i$ on the each physical gauge ``sheet" defined by the ``border" choice of the integration constants. It is important that DST argument of the TWS function $\xi=\frac{1}{\hbar}q_{a}C^{a}, (1\leq a \leq 10)$ will be equal to the action invariant of the single classical material point
\begin{eqnarray}
 S = -a_{\mu}P^{\mu} + \frac{1}{2}\Omega_{\mu \nu} M^{\mu \nu} =const
\end{eqnarray}
under the appropriate choice of these constants. Nevertheless, the solution of the quasi-linear PDE (5.3) is still open.

The DST manifold has only external structure, i.e. only embedding makes the sense for distance in this manifold. I would like to subscribe that this method of coordinatization of DST is in fact the first example outside of general relativity where measurable distance is the function of the energy-momentum.
\section{Conclusion}
I think we should come to term with following facts:

1) A ``body" or even ``elementary" quantum particle cannot be used as primordial elements of the consistent quantum theory. I proposed to use the \emph{unlocated quantum state} (UQS) as the objective ``beable" element of the quantum theory.

2) Our Universe is the functional manifold of quantum amplitudes (like Hilbert space) and the spacetime of our experience is merely some finite dimension gauge ``sheet" of this manifold. Each ``elementary" particle like electron is the superposition of the total quantum states of all Universe in each given spacetime point of the chosen ``sheet".

3) The general rules of the quantum motion hidden in $SU(N)$ group and its sub-manifold
$CP(N-1)$. It means that the \emph{reason of the quantum motion i.e. the existence of the  quantum state} is the coset action of the unitary field.

Quantum Relativity is a new kind of the gauge theory: instead of the adaptation of unitary transformations to spacetime location one needs to accommodate dynamical spacetime structure to the unitary field acting in the space of the UQS's. This gauge theory is state-dependent with variable spacetime structure that should be properly studied.

\end{document}